\documentclass[journal]{IEEEtran}
\usepackage{cite}
\usepackage{graphicx}
\usepackage{amsmath}
\usepackage{times}
\usepackage{latexsym}
\usepackage{graphicx}
\usepackage{bm}
\usepackage{amssymb}
\usepackage[center]{caption2}
\usepackage{stfloats}
\usepackage{cases}
\usepackage{array}
\usepackage{setspace}
\usepackage{fancyhdr}
\usepackage{color}
\usepackage{subfigure}
\usepackage{balance}

\newtheorem{theorem}{Theorem}

\newtheorem{proposition}{Proposition}

\def\proof{\noindent\hspace{2em}{\itshape Proof: }}
\def\endproof{\hspace*{\fill}~$\square$\par\endtrivlist\unskip}

\begin{document}
\title{Optimum Wirelessly Powered Relaying}

\author{Caijun Zhong~\IEEEmembership{Senior Member,~IEEE}, Gan Zheng~\IEEEmembership{Senior Member,~IEEE,} Zhaoyang Zhang~\IEEEmembership{Member,~IEEE}, and George K. Karagiannidis~\IEEEmembership{Fellow,~IEEE}
\thanks{Manuscript received April 06, 2015, accepted April 27, 2015. This work is supported by the National Key Basic Research Program of
China (No. 2012CB316104), the National Natural Science Foundation of
China (61201229), the Zhejiang Science and Technology
Department Public Project (2014C31051), the Zhejiang Provincial Natural
Science Foundation of China (No. LR15F010001). The work of
G. Karagiannidis was supported by the European
Social Fund - ESF and Greek national funds through the Operational Pro-
gram ``Education and Lifelong Learning'' of the National Strategic Reference
Framework (NSRF) - Research Funding Program: THALES-NTUA MIMOSA. The editor coordinating
the review of this paper and approving it for publication was Prof. X. Wang.}
\thanks{C. Zhong and Z. Zhang are with the Institute of Information and Communication Engineering, Zhejiang University, China. (email: caijunzhong@zju.edu.cn).}
\thanks{G. Zheng is with the School of Computer Science \& Electronic Engineering, University of Essex, Wivenhoe Park, Colchester, CO4 3SQ, UK (e-mail: ganzheng@essex.ac.uk).}
\thanks{G. K. Karagiannidis is with Aristotle University of Thessaloniki, Greece, and with Khalifa University, Abu Dhabi, UAE (e-mail: geokarag@ieee.org).}}

\maketitle

\begin{abstract}
This paper maximizes the achievable throughput of a relay-assisted wirelessly powered communications system, where an energy constrained source, helped by an energy constrained relay and both powered by a dedicated power beacon (PB), communicates with a destination. Considering the time splitting approach, the source and relay first harvest energy from the PB, which is equipped with multiple antennas, and then transmits information to the destination. Simple closed-form expressions are derived for the optimal PB energy beamforming vector and time split for energy harvesting and information transmission. Numerical results and simulations demonstrate the superior performance compared with some intuitive benchmark beamforming scheme. Also, it is found that placing the relay at the middle of the source-destination path is no longer optimal.
%
\end{abstract}

\section{Introduction}
Gigabit wireless access will be a reality in fifth generation (5G) wireless systems, with a series of breakthroughs, such as massive multiple-input and multiple-output (MIMO), full-duplex communication and small cell architectures, which in turn, has fueled a number of emerging wireless services such as mobile gaming, mobile TV and mobile Internet. With the proliferation of smartphones and tablets, one of the most critical issues affecting the user experience is the limited operation lifetime of mobile devices due to finite battery capacity. Motivated by this, radio-frequency (RF) based energy harvesting technique has received substantial research interests in recent years \cite{B.Medepally,W.Lumpkins}. Empowered with the RF energy harvesting capability, it is possible to virtually provide perpetual energy supply to mobile devices, eliminating the need to plug into the power grid for battery recharging.

RF energy harvesting, combined with information transfer, has resulted in a new emerging topic, namely, simultaneous wireless information and power transfer (SWIPT), which has attracted significant attention from academia and recently from industry. Thus far, various aspects of SWIPT systems have been investigated, including information theoretical limits \cite{L.Varshney,P.Grover}, practical architectures \cite{R.Zhang,L.Liu,X.Zhou}, effect of imperfect channel state information \cite{M.Tao}, OFDM-based SWIPT systems \cite{Derrick,K.Huang3} and relay-assisted SWIPT systems \cite{A.Nasir,Z.Ding,Z.Ding1,Z.Ding2,C.Zhong,I.Krikidis_2,I.Krikidis}. It is worth pointing out here that all these prior works assume that the mobile devices harvest energy from ambient RF signals. However, even if this may be viable for low power devices such as sensors, it is in general infeasible to power larger devices such as smartphones, tablets and laptops \cite{K.Huang1}. Responding to this key limitation, the authors in \cite {K.Huang2} proposed a novel network architecture, where cellular base stations are underlaid by dedicated power beacons (PB), which can be used to supply energy to mobile devices through microwave power transfer. Since the PBs do not require any backhaul link, the associated cost of PB deployment is much lower, hence, dense deployment of PBs to ensure network coverage for a wide range of mobile devices is feasible.

In this paper, we consider a dual-hop decode-and-forward (DF) relaying system, where both the source and relay are powered by the dedicated PB. To improve the energy transfer efficiency, the multiple antenna enabled PB performs energy beamforming. It is assumed that wireless power transfer is performed over the same frequency as information transfer, and the time-switching protocol \cite{R.Zhang} is adopted. As such, the single antenna at the source and relay switches between two hardware chains, one for energy harvesting and one for information transmission/reception. Therefore, the entire communication block consists of two different phases, i.e., energy harvesting phase, where the source and relay harvest the energy from the PB, and information transfer phase, where the relay assists the information transmission between the source and destination.

The main contribution of this paper is the derivation of simple closed-form expressions for the optimal time split between the energy harvesting and information transfer phase, as well as the optimal energy beamforming vectors at the PB, which maximize the system's throughput. Simulation results demonstrate that the optimal solution substantially outperforms the intuitive benchmark beamformer. In addition, it is revealed that placing the relay in the middle of the source and destination path is no longer optimal, instead, the distances to the PB should also be taken into consideration when optimizing the relay position.

{\it Notations}: $(\cdot)^*$ denotes the complex conjugate, $(\cdot)^T$ denotes the matrix transpose, and $(\cdot)^{\dag}$ denotes the Hermitian transpose. ${\bf I}$ is the identity matrix of appropriate size. $\Pi_{{\bf X}} = {\bf X}({\bf X}^{\dag}{\bf X})^{-1}{\bf X}^{\dag}$ is the orthogonal projection onto the column space of ${\bf X}$, and $\Pi_{{\bf X}}^{\perp}={\bf I}-\Pi_{{\bf X}}$ is the orthogonal projection onto the orthogonal complement of the column space of ${\bf X}$.

\section{System Model}
We consider a communication system where the source S communicates with destination D with the assistance of the relay R, as depicted in Fig. \ref{fig:fig0}. We assume that both S and R are energy constrained, hence rely on the external energy charging via wireless power transfer from a dedicated PB. All three communication nodes are equipped with a single antenna, while the PB is equipped with $N$ antennas. Full channel state information (CSI) of the PB and source and relay links is assumed at the PB. In practice, the channel can be estimated by overhearing the pilot send by the source and relay. In addition, the channel magnitudes of the two hop information links are assumed to be known at the PB, which, for instance, could be obtained through feedback from the relay.

\begin{figure}[htb!]
\centering
\includegraphics[clip,viewport=100 360 600 650,scale=0.38]{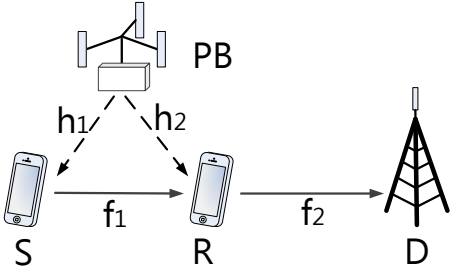}
\caption{System model}\label{fig:fig0}
\end{figure}


The entire communication consists of two different phases, namely, {\it energy harvesting} and {\it information transmission} phase. Assuming a block time of $T$, during the first phase of duration $\tau T$, where $0<\tau<1$, S and R harvest energy from the PB. The remaining time of duration $(1-\tau)T$ is equally partitioned into two parts, during the first half period, S transmits information to R and during the second half, R forwards the information to D.

During the energy harvesting phase, the received signal at S and R can be expressed, respectively, as
\begin{align}
y_s^e &= \sqrt{Pd_1^{-\alpha}}{\bf h}^T_1{\bf x}_e+n_s\mbox{ and } y_r^e &= \sqrt{Pd_2^{-\alpha}}{\bf h}^T_2{\bf x}_e+n_r,\notag
\end{align}

where $P$ is the transmit power at the PB, $d_1$ and $d_2$ denote the distances between PB and S, PB and R, respectively. Furthermore, $\alpha$ is the path loss exponent, ${\bf h}_1$ and ${\bf h}_2$ are the channel vectors of size $N\times 1$, ${\bf x}_e$ is an $N\times 1$ vector satisfying ${\tt E}\left\{{\bf x}_e^\dag{\bf x}_e\right\}=1$, and $n_s$ and $n_r$ are the zero-mean additive white Gaussian noise (AWGN) samples with variance $N_0$.

Since the PB is equipped with multiple antennas, energy beamforming could be applied to improve the efficiency of energy transfer, i.e., ${\bf x}_e={\bf w}s_e$,
where ${\bf w}$ is the beamforming vector with $\|{\bf w}\|^2 =1$, while $s_e$ is the energy symbol with unit power. As such, the total received energy at the S and R at the end of the first phase can be expressed as
\begin{equation}
E_s = \eta|{\bf w}^T{\bf h}_1|^2P\tau Td_1^{-\alpha},
\end{equation}
and
\begin{equation}
 E_r = \eta|{\bf w}^T{\bf h}_2|^2P\tau Td_2^{-\alpha},
\end{equation}
respectively, where $0<\eta<1$ is the energy conversion efficiency.

In the first half of the second phase, S transmits information to R using the energy harvested in the first phase. Hence, the received signal $y_r$ at R is given by
\begin{align}
y_r=\sqrt{\frac{2E_s}{(1-\tau)Td_3^{\alpha}}}f_1s_0+n_r,
\end{align}
where $d_3$ denotes the distance between S and R, $f_1$ is the channel coefficient, $s_0$ is the information symbol with unit energy.

Upon receiving the source signal, the relay first decodes the source symbol and then forwards it to the destination using the energy harvested during the first phase,\footnote{Please note, we have ignored the processing power required by the transmit/receive circuitry at the relay as in \cite{A.Nasir,X.Zhou,B.Med}. This assumption is justifiable since the transmission energy is the dominant source of energy consumption.} as such the signal at D can be expressed as
\begin{align}
y_d=\sqrt{\frac{2E_r}{(1-\tau)Td_4^{\alpha}}}f_2s_r+n_d,
\end{align}
where $d_4$ denotes the distance between R and D, $f_2$ is the channel coefficient, $s_r$ is the relay signal with unit energy, and $n_d$ is a sample of the AWGN with variance $N_0$.

Therefore, the effective end-to-end SNR at the destination can be written as
\begin{align}
\gamma
&=\frac{2\tau\eta P}{(1-\tau)N_0}\min\left\{\frac{|{\bf w}^T{\bf h}_1|^2|f_1|^2}{d_1^{\alpha}d_3^{\alpha}}, \frac{|{\bf w}^T{\bf h}_2|^2|f_2|^2}{d_2^{\alpha}d_4^{\alpha}}\right\}.\label{eqn:e2esnr}
\end{align}

%

\section{Throughput Optimization}
The achievable system's throughput, $R$, can be expressed as
\begin{align}
&R = \frac{(1-\tau)}{2}\times\\
&\log_2\left(1+\frac{2\tau\eta P}{(1-\tau)N_0}\min\left\{\frac{|{\bf w}^T{\bf h}_1|^2|f_1|^2}{d_1^{\alpha}d_3^{\alpha}}, \frac{|{\bf w}^T{\bf h}_2|^2|f_2|^2}{d_2^{\alpha}d_4^{\alpha}}\right\}\right).\notag
\end{align}
Hence, the following optimization problem is created:
\begin{align}
{\cal P}1:\quad \max_{\tau, {\bf w}} & \quad R\\
\mbox{s.t.} & \quad 0<\tau <1\\
&\quad \|{\bf w}\|^2 =1.
\end{align}
At the first glance, the above problem requires the joint optimization of $\tau$ and ${\bf w}$, which is in general difficult. Nevertheless, a close observation reveals that the special structure of ${\cal P}1$ allows for a separate optimization of $\tau$ and ${\bf w}$. Specifically, we present the following key result.
\begin{proposition}
The original optimization problem, ${\cal P}1$, is equivalent to the following:
\begin{align}
{\cal P}2:\quad \max_{\tau} & \quad  \frac{(1-\tau)}{2}\log_2\left(1+\frac{2\tau\eta P}{(1-\tau)N_0}z_{m}\right)\\
\mbox{s.t.} & \quad 0<\tau <1,
\end{align}
where $z_{m}$ is defined as
\begin{align}
z_{m}=\min\left\{\frac{|\hat{\bf w}^T{\bf h}_1|^2|f_1|^2}{d_1^{\alpha}d_3^{\alpha}}, \frac{|\hat{\bf w}^T{\bf h}_2|^2|f_2|^2}{d_2^{\alpha}d_4^{\alpha}}\right\},
\end{align}
with $\hat{\bf w}$ being the solution of the following optimization problem:
\begin{align}
{\cal P}3:\quad \max_{\bf w} &\quad \min\left\{\frac{|{\bf w}^T{\bf h}_1|^2|f_1|^2}{d_1^{\alpha}d_3^{\alpha}}, \frac{|{\bf w}^T{\bf h}_2|^2|f_2|^2}{d_2^{\alpha}d_4^{\alpha}}\right\}\\
\mbox{s.t.} &\quad \|{\bf w}\|^2 =1.
\end{align}
\end{proposition}
\proof
Define function  $g(\tau,z)$ as
\begin{align}
g(\tau,z) = \frac{(1-\tau)}{2}\log_2\left(1+\frac{2\tau\eta P}{(1-\tau)N_0}z\right),
\end{align}
where $0<\tau <1$ and $z$ is a positive real number. It is easy to prove that function $g(\tau,z)$ is an increasing function with respect to $z$. Now consider two positive real numbers $z_1$ and $z_2$ such that $z_2>z_1$, and let $\tau_i$ being the value of $\tau$ which maximizes $g(\tau,z_i)$, i.e., $g(\tau_i,z_i)\geq g(\tau,z_i)$ for all $\tau$, $i=1, 2$. Then, it holds that
\begin{align}
g(\tau_2,z_2)\geq g(\tau_1,z_2)\geq g(\tau_1,z_1),
\end{align}
which indicates that the maximum of $g(\tau,z)$ is achieved at the point when $z$ attains its maximum. Therefore, a sequential optimization of problems ${\cal P}3$ and ${\cal P}2$ yields the optimal solution for the original problem ${\cal P}1$.
\endproof

In the following, we investigate the optimal solutions for the problems ${\cal P}2$ and ${\cal P}3$.
\begin{proposition}
The optimal $\tau$ for the optimization problem ${\cal P}2$ is given by
\begin{align}
\hat{\tau} = \frac{e^{W\left(\frac{\beta-1}{e}\right)+1}-1}{\beta+e^{W\left(\frac{\beta-1}{e}\right)+1}-1},
\end{align}
where $W(x)$ is the Lambert W function \cite{W}, and $\beta = \frac{2\eta P z_m}{N_0}$.
\end{proposition}
\proof The proof follows from the results presented in \cite[Appendix A]{C.Zhong}.\endproof

We now turn to problems ${\cal P}3$, and we have the following key result:

\begin{theorem}\label{theorem:1}
The optimal beamforming vector $\hat{\bf w}$ for the optimization problem ${\cal P}3$ can be expressed as
\begin{align}
\hat{\bf w} = \bar{x} \frac{\Pi_{\hat{\bf h}_2^*}\hat{\bf h}_1^*}{\|\Pi_{\hat{\bf h}_2}\hat{\bf h}_1\|}+\sqrt{1-\bar{x}^2} \frac{\Pi_{\hat{\bf h}_2^*}^{\perp}\hat{\bf h}_1^*}{\|\Pi_{\hat{\bf h}_2}^{\perp}\hat{\bf h}_1\|},
\end{align}
where $\hat{{\bf h}}_1=\frac{|f_1|}{\sqrt{{d_1^{\alpha}d_3^{\alpha}}}}{\bf h}_1$, $\hat{{\bf h}}_2=\frac{|f_2|}{\sqrt{{d_2^{\alpha}d_4^{\alpha}}}}{\bf h}_2$, $a = \|\Pi_{\hat{\bf h}_2}\hat{\bf h}_1\|$, $b=\|\Pi_{\hat{\bf h}_2}^{\perp}\hat{\bf h}_1\|$, and $c= \frac{|\hat{\bf h}_1^{\dag}\Pi_{\hat{\bf h}_2}\hat{\bf h}_2|}{\|\Pi_{\hat{\bf h}_2}\hat{\bf h}_1\|}$, and $\bar{x}$ being
\begin{align}
\bar{x} = \left\{\begin{array}{cc}
                \frac{a}{\sqrt{a^2+b^2}},& c\geq \frac{a^2+b^2}{a},\\
                \frac{b}{\sqrt{(a-c)^2+b^2}},&a \leq c< \frac{a^2+b^2}{a},\\
                1,&c<a.\\
                \end{array}
                \right.
\end{align}

\end{theorem}

\proof
According to \cite[Corollary 1]{E.Larsson}, the optimal beamforming vector $\hat{\bf w}$ can be expressed as
\begin{align}
\hat{\bf w} = x \frac{\Pi_{\hat{\bf h}_2^*}\hat{\bf h}_1^*}{\|\Pi_{\hat{\bf h}_2}\hat{\bf h}_1\|}+\sqrt{1-x^2} \frac{\Pi_{\hat{\bf h}_2^*}^{\perp}\hat{\bf h}_1^*}{\|\Pi_{\hat{\bf h}_2}^{\perp}\hat{\bf h}_1\|},
\end{align}
where $x\in[0,1]$. Now let us define $g_1(x)\triangleq|\hat{\bf w}^T\hat{\bf h}_1| =a x +b\sqrt{1-x^2}$, and $g_2(x)\triangleq |\hat{\bf w}^T\hat{\bf h}_2|=cx$. Then, the original optimization problem ${\cal P}1$ is equivalent to maximize the function $g(x)$, i.e.,
\begin{align}
\max_{x}&\quad g(x),\quad\quad \mbox{s.t. }\quad 0\leq x\leq 1,
\end{align}
where $g(x)$ is defined as
\begin{align}
g(x) \triangleq \min\left(g_1(x), g_2(x)\right)=\min\left(ax+b\sqrt{1-x^2}, cx\right).\notag
\end{align}
It is easy to show that $g_1(x)$ is a concave function with respect to $x$, hence, its maximum can be attained by solving $g'_1(\hat{x}) = 0$, which gives $\hat{x}=\frac{a}{\sqrt{a^2+b^2}}$, and $g_1(\hat{x})=\sqrt{a^2+b^2}$.

\begin{figure}[htb]
  \centering
  \subfigure[Case 1]{
    \label{fig:subfig:a} 
    \includegraphics[width=1.5in]{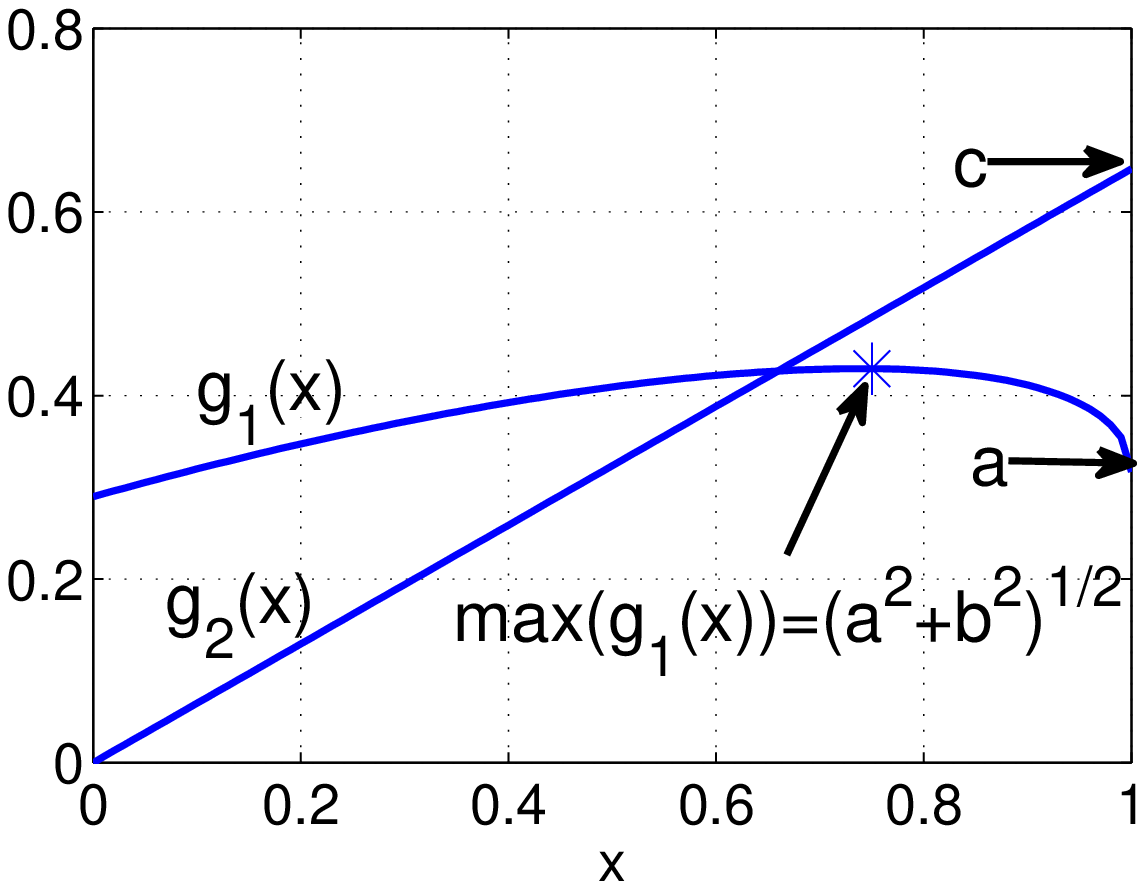}}
  \subfigure[Case 2]{
    \label{fig:subfig:b} 
    \includegraphics[width=1.5in]{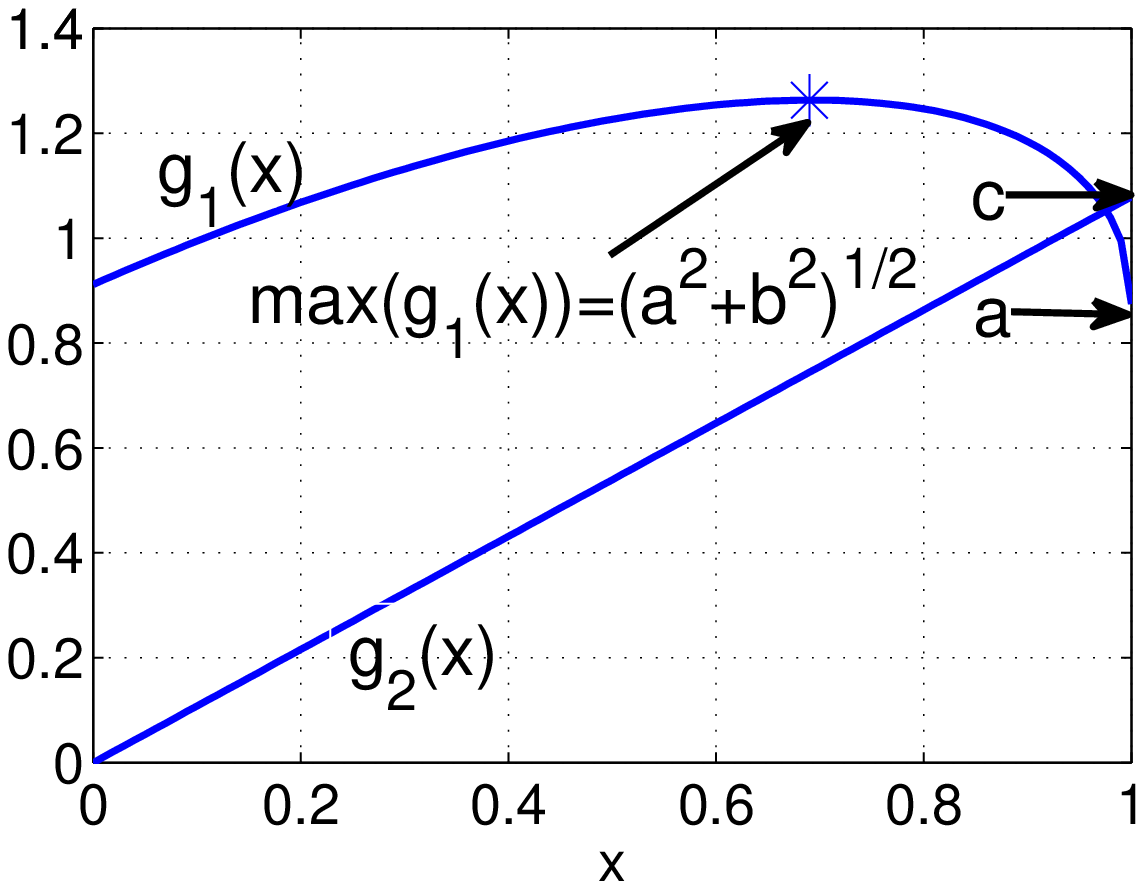}}
  \subfigure[Case 3]{
    \label{fig:subfig:c} 
    \includegraphics[width=1.5in]{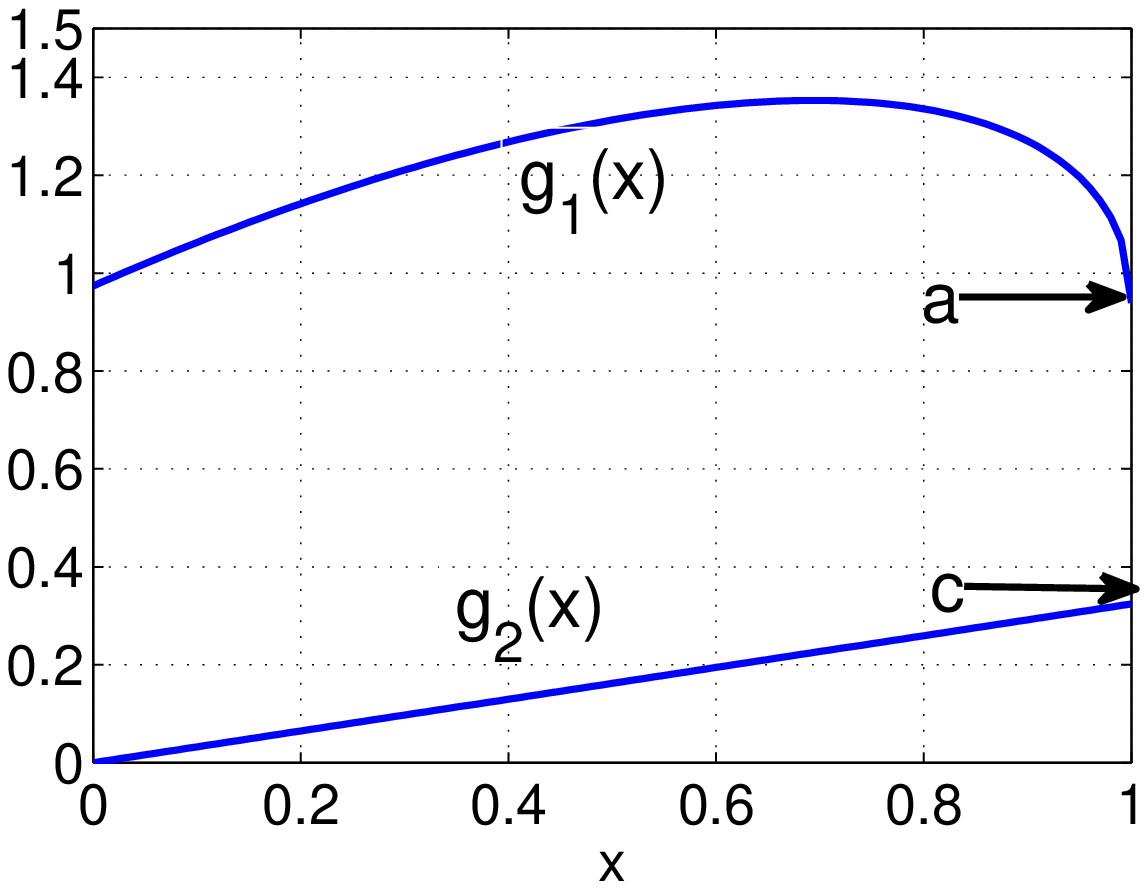}}
  \caption{Three different cases for the maximization of $g(x)$.}
  \label{fig:subfig} 
\end{figure}

Now, as shown in Fig. \ref{fig:subfig}, we have three different scenarios:
\begin{itemize}
\item In Case 1, we observe is that the x-label of the cross point of $g_1(x)$ and $g_2(x)$, i.e., $x_c$, can be characterized by $g_1(x_c)=g_2(x_c)$, which gives $x_c = \frac{b}{\sqrt{b^2+(c-a)^2}}$. If the slope of $g_2(x)$ is sufficiently large, such that the cross point appears before $g_1(x)$ attains its maximum. i.e., $x_c\leq \hat{x}$, which gives $c\geq \frac{a^2+b^2}{a}$, then, the maximum of $g(x)$ is achieved at $\bar{x}=\hat{x} = \frac{a}{\sqrt{a^2+b^2}}$, which is also the maximum point of $g_1(x)$, as shown in Fig. \ref{fig:subfig:a}.

\item In Case 2, the cross point appears after $g_1(x)$ attains its maximum, i.e., $a \leq c< \frac{a^2+b^2}{a}$, then, the maximum of $g(x)$ is achieved at the cross point, i.e., $\bar{x}=x_c=\frac{b}{\sqrt{(a-c)^2+b^2}}$, as shown in Fig. \ref{fig:subfig:b}.
\item In Case 3, there is no cross point between $g_1(x)$ and $g_2(x)$, i.e., $g_1(x)\geq g_2(x)$ for all $x$, namely, $c< a$, then, the maximum of $g(x)$ is  identical to the maximum of $g_2(x)$ which is achieved at the point $\bar{x}=1$, as shown in Fig. \ref{fig:subfig:c}.
\end{itemize}

\endproof

\section{Numerical Results and Discussion}
In this section, numerical and simulations results are presented to illustrate the impact of key system parameters on the system's throughput. Without loss of generality, we set the energy conversion efficiency $\eta = 0.4$, and path-loss exponent $\alpha=3$. Please note, the throughput is obtained by averaging over $10^3$ independent channel realizations.

To demonstrate the superiority of the proposed optimal scheme, we compare it with an intuitive benchmark scheme by looking into the asymptotic large antenna regime, where the optimal beamforming vector becomes
\begin{align}
&\bar{\bf w} =\left(\frac{|f_2|}{\sqrt{d_2^\alpha d_4^\alpha}}\frac{{\bf h}_1^*}{\|{\bf h}_1\|} +\frac{|f_1|}{\sqrt{d_1^\alpha d_3^\alpha}}\frac{{\bf h}_2^*}{\|{\bf h}_2\|}\right)\left/\sqrt{\frac{|f_2|^2}{{d_2^\alpha d_4^\alpha}} +\frac{|f_1|^2}{{d_1^\alpha d_3^\alpha}}}\right..
\end{align}
The rationale behind the choice of $\bar{\bf w}$ is that, the optimal beamforming vector should be a linear combination of ${\bf h}_1^*$ and ${\bf h}_2^*$, hence, the key is to design the optimal weights. Capitalizing on the asymptotical orthogonality of ${\bf h}_1^*$ and ${\bf h}_2^*$ when $N\rightarrow \infty$, the optimal weights can be easily obtained.

Fig. \ref{fig:fig0} depicts the achievable throughput of the optimal scheme and the benchmark scheme with $d_1=d_2=3\mbox{m}$ and $d_3=d_4=5\mbox{m}$. It can be readily observed that the optimal scheme outperforms the benchmark scheme, and the performance gap is rather significant for moderate number of antennas $N$. On the other hand, when $N$ is sufficiently large, i.e., $N=5000$, the performance gap narrows substantially. This is rather expected since the benchmark scheme becomes asymptotically optimal. We also observe that the throughput improves when $N$ increases, which is also intuitive since the energy transfer efficiency improves with a large size of antenna array.
\begin{figure}[htb!]
\centering
\includegraphics[scale=0.43]{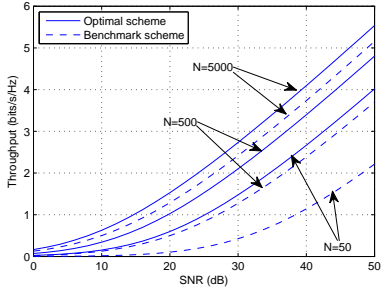}
\caption{Throughput comparison between the optimal scheme and benchmark scheme for different $N$.}\label{fig:fig0}
\end{figure}
%
%
%

\begin{figure}[htb!]
\centering
\includegraphics[scale=0.43]{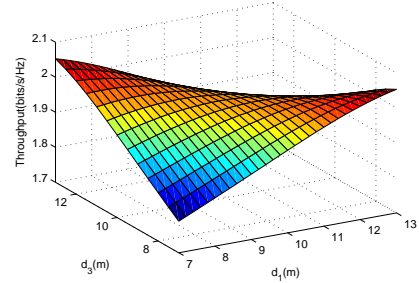}
\caption{Impact of node positions on the system throughput.}\label{fig:fig1}
\end{figure}

Fig. \ref{fig:fig1} examines the impact of node positions on the throughput performance when $N=10$, $d_1+d_2=20\mbox{m}$, and $d_3+d_4=20\mbox{m}$, as a function of $d_1$ and $d_3$, both vary from $7$ to $13$. It can be observed that higher throughput is attained at the point $d_1=7$ and $d_3 = 13$, a scenario where the PB is close to the source while the relay is close to the destination; or at the point $d_1 =13$ and $d_3 = 7$, a scenario where the PB is close to the relay while the relay is close to the source. The above results are somehow intuitive, since the performance of dual-hop relaying systems is bottlenecked by the weakest link, hence, an optimized system shall achieve a fine balance between the two hops.

Another interesting observation from Fig. \ref{fig:fig1} is that the symmetric setup, i.e., the point $d_1=d_2=10$ and $d_3=d_4=10$, does not yield the maximum throughput. This is in sharp contrast to the conventional dual-hop relaying systems where it is always desirable to put the relay node in the middle of the source and destination link. The reason is that, with the introduction of PB, in addition to the distance of the information transfer links, the throughput performance also heavily depends on the distance of the energy transfer links. As a matter of fact, the throughput is determined by $d_1^\alpha d_3^\alpha$ as shown in the end-to-end SNR expression (\ref{eqn:e2esnr}). Since $(7\times 13)^\alpha <(10\times 10)^\alpha$, it becomes obvious why the maximum throughput is achieved at point $d_1=3$ and $d_3 = 13$.

%
%

\begin{figure}[htb!]
\centering
\includegraphics[scale=0.43]{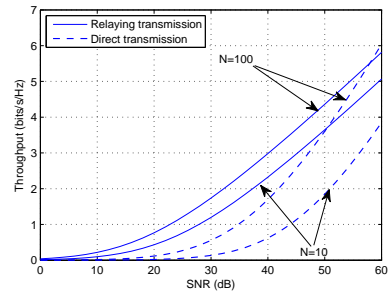}
\caption{Throughput comparison: Relaying v.s. Direct transmission.}\label{fig:fig00}
\end{figure}

Fig. \ref{fig:fig00} compares the achievable throughput of the dual-hop relaying system with that of direct transmission system with optimized time split and beamforming vector. As can be readily observed, at the SNR levels of practical interest, i.e., $0\mbox{ dB}<\mbox{SNR}< 50 \mbox{ dB}$, adopting the relaying structure improves the system throughput. Moreover, the performance gap is more substantial with moderate number of antennas, i.e., $N=10$. Only if the transmit SNR is very high, direct transmission becomes preferred. This is rather intuitive since at the high SNR regime, the system is degree-of-freedom limited, as such the half-duplex relay operation becomes the bottleneck as manifested through the $1/2$ factor in the throughput expression. Since in the wirelessly powered communications systems, the source is likely to operate in the power-limited regime, adopting the relaying structure is beneficial in general.

\section{Conclusion}
In this paper, we have optimized the throughput of a relay-assisted wirelessly powered communication system. Specifically, we obtained simple closed-form solutions for the PB energy beamforming vector as well as the optimal time split for the energy harvesting phase and information transmission phase. It was shown that the optimal solution yields significant performance gain compared to the intuitive benchmark scheme.

\end{document}